\documentclass[aps,prl,reprint,amsmath,amssymb,amsfontsprl,superscriptaddress
]{revtex4-2}
\usepackage[colorlinks,
            linkcolor=blue,
            anchorcolor=blue,  
            citecolor=blue,
            urlcolor=blue
            ]{hyperref}
\usepackage{graphicx}
\usepackage{bm}
\usepackage{color}
\usepackage{float}
\begin{document}

\title{Non-Hermitian singularities induced single-mode depletion and soliton formation in microresonators}

\author{Boqing Zhang}
\thanks{These authors contributed equally to this work.}
\author{Nuo Chen}
\thanks{These authors contributed equally to this work.}
\author{Haofan Yang}
\author{Yuntian Chen}
\email{yuntian@hust.edu.cn}
\affiliation{%
    School of Optical and Electronic Information, Huazhong University of Science and Technology, Wuhan 430074, China
}%
\affiliation{%
    Wuhan National Laboratory for Optoelectronics, Huazhong University of Science and Technology, Wuhan 430074, China
}%
\author{Heng Zhou}
\affiliation{%
    Key Lab of Optical Fiber Sensing and Communication Networks, University of Electronic Science and Technology of China, Chengdu 611731, China
}%
\author{Xinliang Zhang}
\email{xlzhang@hust.edu.cn}
\author{Jing Xu}
\email{jing\_xu@hust.edu.cn}
\affiliation{%
    School of Optical and Electronic Information, Huazhong University of Science and Technology, Wuhan 430074, China
}%
\affiliation{%
    Wuhan National Laboratory for Optoelectronics, Huazhong University of Science and Technology, Wuhan 430074, China
}%

\date{\today}

\begin{abstract}
    On-chip manipulation of single resonance over broad background comb spectra of microring resonators is indispensable, ranging from tailoring laser emission, optical signal processing to non-classical light generation, yet challenging without scarifying the quality factor or inducing additional dispersive effects. Here, we propose an experimentally feasible platform to realize on-chip selective depletion of single resonance in microring with decoupled dispersion and dissipation, which are usually entangled by Kramer-Kroning relation. Thanks to the existence of non-Hermitian singularity, unsplit but significantly increased dissipation of the selected resonance is achieved due to the simultaneous collapse of eigenvalues and eigenvectors, fitting elegantly the requirement of pure mode depletion. With delicate yet experimentally feasible parameters, we show explicit evidence of modulation instability as well as deterministic single soliton generation in microresonators induced by depletion in normal and anomalous dispersion regime, respectively. Our findings connect non-Hermitian singularities to wide range of applications associated with selective single mode manipulation in microwave photonics, quantum optics, ultrafast optics and beyond.
\end{abstract}

\maketitle

Selective manipulation of cavity resonant modes including amplification, depletion, linewidth modification etc. has far reaching implications and applications in optics \cite{Feng972,Hodaei975,9205209,noda_spontaneous-emission_2007,PhysRevLett.116.061102}. In particular, as the rapid development of nanofabrication technology, miniaturized ring resonators have matured to be an ideal platform for the study of nonlinear optics \cite{shen_integrated_2020}. Mode manipulation of such micro-cavities, especially on-demand spectrum engineering, plays a crucial role in a vast of nonlinear applications, including optical signal processing \cite{morichetti_travelling-wave_2011,8423644}, non-classical light generation \cite{heuck_unidirectional_2019,zhang_squeezed_2021,zeng_four-wave_2015,liu_high-spectral-purity_2020}, optical frequency comb generation \cite{xue_mode-locked_2015,bao_spatial_2017,liao_enhanced_2017,xue_super-efficient_2019,pasquazi_micro-combs_2018,herr_mode_2014,yi_single-mode_2017}.

Notably, mode manipulations of microring resonators are routinely realized by Hermitian mode coupling with either high-order spatial modes \cite{xue_mode-locked_2015,bao_spatial_2017,herr_mode_2014} or auxiliary resonators \cite{heuck_unidirectional_2019,zhang_squeezed_2021,xue_normal-dispersion_2015,gentry_tunable_2014}. This is due to the fact that non-touching scheme based on mode coupling is an excellent option to avoid scarifying the quality factor of microring resonators, which is of vital importance in most nonlinear applications. Essentially, mode-coupling based mode manipulation gives rise to dispersion effect due to resonance splitting, which is exceptionally relevant to the phase matching condition of optical parametric process. On the other hand, mode-coupling based mode manipulation fails to realize a pure mode-selective depletion without dispersion effect, as fundamentally limited by Kramer-Kroning relation. Indeed, auxiliary microresonator is only used as a dispersion tool \cite{xue_normal-dispersion_2015} to realize modulation instability (MI) in normal group-velocity dispersion (GVD) regime. In contrast, pure mode-selective dissipation \cite{PhysRevLett.93.163902} have been experimentally implemented in fiber-based links \cite{PhysRevLett.93.163902,8868166} or loops \cite{bessin_gain-through-filtering_2019} to realize MI and frequency comb in normal GVD regime, but absent in microring resonators due to the entanglement between dispersion and dissipation dictated by Kramer-Kroning relation. Therefore, it is important to identify alternative strategies to isolate the impact of depletion from dispersion.

In this work, non-Hermitian mode coupling is investigated and the exotic properties of non-Hermitian singularity are explored to selectively deplete target resonance of microring resonators with minimized dispersion effects. We note that a number of seminar papers have studied the coupled microring resonators with the symmetric structure, operated in PT unbroken and/or broken phase \cite{Hodaei975,Peng328,milian_cavity_2018,komagata_dissipative_2021}, where either the single mode selectivity is absent \cite{milian_cavity_2018,komagata_dissipative_2021} or selectivity of single mode location depends on the gain spectrum \cite{Hodaei975,Peng328}. Meanwhile, non-Hermitian singularities are explored to enhance sensitivity \cite{hodaei_enhanced_2017} or Sagnac effect \cite{lai_observation_2019,hokmabadi_non-hermitian_2019}. In contrast, the underlying principle of our scheme is to explore coupled ring resonators with different radius and take the advantage of simultaneous collapse of eigenvalues and eigenvectors at exceptional point (EP). In this scenario, the line-shape functions of the coupled resonators at target frequency are identically overlapped where the quality (Q)-factor is squeezed to its minimum, offering non-Hermitian singularity induced dispersion-free depletion. Based on such effect, dissipation induced MI are predicted in microresonator in normal GVD regime, resulting in the generation of optical Kerr frequency comb as well as dark solitons. In addition, single-mode depletion is shown to greatly improve the success rate of single soliton generation in the anomalous dispersion regime. 
\begin{figure}[htbp]
    \centering
    \includegraphics[scale=0.585]{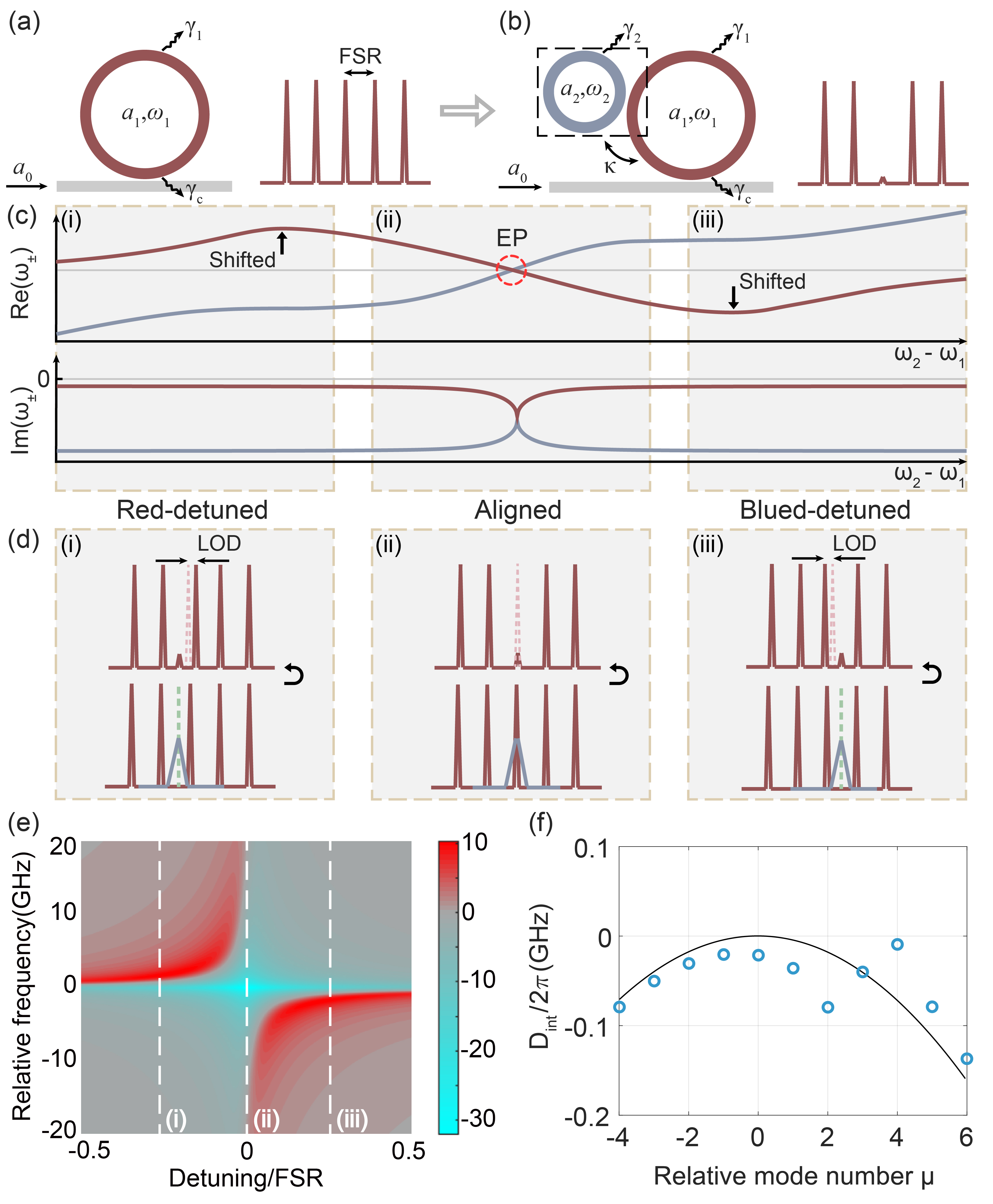}
    \caption{Schematic diagram of (a) a conventional high-Q microresonator and (b) parity-time symmetric coupled microrings and their corresponding intracavity field enhancement of the main cavity. (c) Evolution of $Re(\omega_{\pm})$ and $Im(\omega_{\pm})$ as a function of detuning ($\Delta$). (d) Typical intracavity field enhancement at $\Delta<0$ (red-detuned, i), $\Delta=0$ (aligned, ii) and $\Delta>0$ (blue-detuned, iii). (e) $|\eta(\omega)|^2$ as a funtion of $\delta$. (f) Local dispersion (LOD) with (blue circles) and without (black line) mode coupling at $\Delta=0$. Parameters used in (e),(f) are $D_1/2\pi\approx226$GHz, $D_2/2\pi\approx-8.65$MHz (the second order dispersion parameter), $\gamma_1\approx271.65$MHz, $\gamma_c\approx42.45$MHz, $\gamma_c\approx56$GHz, and $\kappa\approx23.4$GHz. The circumference ratio of the main cavity and the auxiliary cavity is 33.}
    \label{p1}
\end{figure}

To illustrate the idea of single mode depletion, we compare the case of non-Hermitian coupled resonators with a single microring resonator, both of which are coupled to a straight waveguide (Fig. 1). It is well-known that the intracavity field enhancement (FE) of the single resonator, defined as the intracavity mode field $a_1$ divided by input mode field $a_0$, yields a comb spectrum (Fig. 1(a)). In parallel, Fig. 1(b) sketches a passive parity-time symmetrically coupled structure, wherein the main ring resonator (red, high-Q) at $\omega_1$ is coupled to a thermally-tunable auxiliary ring resonator (blue, low-Q, intracavity mode field $a_2$) at $\omega_2$ with coupling coefficient $\kappa$. The intrinsic decay rates of two resonator are denoted by $\gamma_1$ and $\gamma_2$, respectively, with the coupling decay rate $\gamma_c$ for the main ring. The radius of the auxiliary ring is much smaller than the main ring. In the regime of $\gamma_1+\gamma_c\ll\gamma_2$, the evolution of modal field $A$=($a_1,a_2$) in this structure obeys $idA/dt=H\!A$, where non-Hermitian Hamiltonian $H$ is given as follows
\begin{equation}\label{equation1}
    H=
    \begin{pmatrix}
    \omega_1 & \kappa \\
    \kappa & \omega_2-i\gamma_2
    \end{pmatrix}
\end{equation}
The eigenfrequencies $\omega_\pm=\frac{1}{2}\left[\left(\omega_1+\omega_2\right)-i\gamma_2\right]\pm\frac{1}{2}\sqrt{4\kappa^2-[\gamma_2+i(\omega_1-\omega_2)]^2}$ are plotted schematically in Fig. 1(c) as a function of frequency detuning $\Delta=\omega_2-\omega_1$. Accordingly, the schematic FE of the main resonator exhibit dispersion effects in the case of detuning ($\Delta\!\neq0$, Figs. 1(d)(i) and (iii)). In contrast, FE of zero detuning case ($\Delta=0$)  features a pure depletion effect (Fig. 1(d)(ii)). The dispersion-free depletion can be described quantitatively by the ratio $\eta(\omega)$ between FE of the case shown in Fig. 1(b), i.e., $a_1/a_0$ and the counterpart shown in Fig. 1(a). From the coupled mode analysis, $\eta(\omega)$ is found to be
\begin{equation}\label{e1}
  \eta(\omega)=1+\frac{\kappa^2}{(\omega-\omega_+)(\omega-\omega_-)}
\end{equation}
Under the conditions of $4\kappa^2=\gamma_2^2$ and $\Delta=0$, eigenfrequencies and eigenvectors collapses and the coupled resonators operates at EP. It is trivial to see that $\omega_\pm\approx\omega_1-i\gamma_2/2$ and $\eta\left(\omega_1\right)=0$ at EP, leading to the absence of resonance split (dispersion-free) and elimination of resonant field at $\omega_1$ (depletion). Using practical parameters, Fig.1(e) shows the colour-map of $\left|\eta\left(\omega\right)\right|^2$ as a function of detuning, which is fully consistent with Fig. 1(d). Figure 1(f) shows the local dispersion (LOD) in case of zero detuning with (blue circles) and without (black line) non-Hermitian mode coupling. LOD is defined as $D_{int}\left(\mu\right)=\omega_\mu-\omega_0-D_1\mu$, where $D_1$ and $\mu$ correspond to the free spectral range (FSR) of the main ring and the relative mode index of the main cavity with respect to the pump (designated by $\mu=0$)\cite{herr_mode_2014}. LOD vanishes at $\mu=3$, verifying the dispersion and dispersion-free depletion effects shown in Fig. 1(d)(ii). The residual dispersion observed at $\mu\neq3$ arises from limited FSR. To be specific, the auxiliary resonance that is aligned with $\mu=3$ is misaligned to other modes of the main cavity, resulting similar effect as the case of $\Delta\neq0$. Nevertheless, such effect is weak due to large frequency misaligned and is shown to have limited impact in the latter frequency comb analysis. In the rest of this paper, the coupled ring resonators operates at $4\kappa^2=\gamma_2^2$ without otherwise specified, which can be satisfied by tuning the coupling $\kappa$ with micro-heaters \cite{gentry_tunable_2014} or $\gamma_2$ with tunable loss mechanism \cite{Peng328}.

\begin{figure}[htbp]
    \centering
    \includegraphics[scale=0.53]{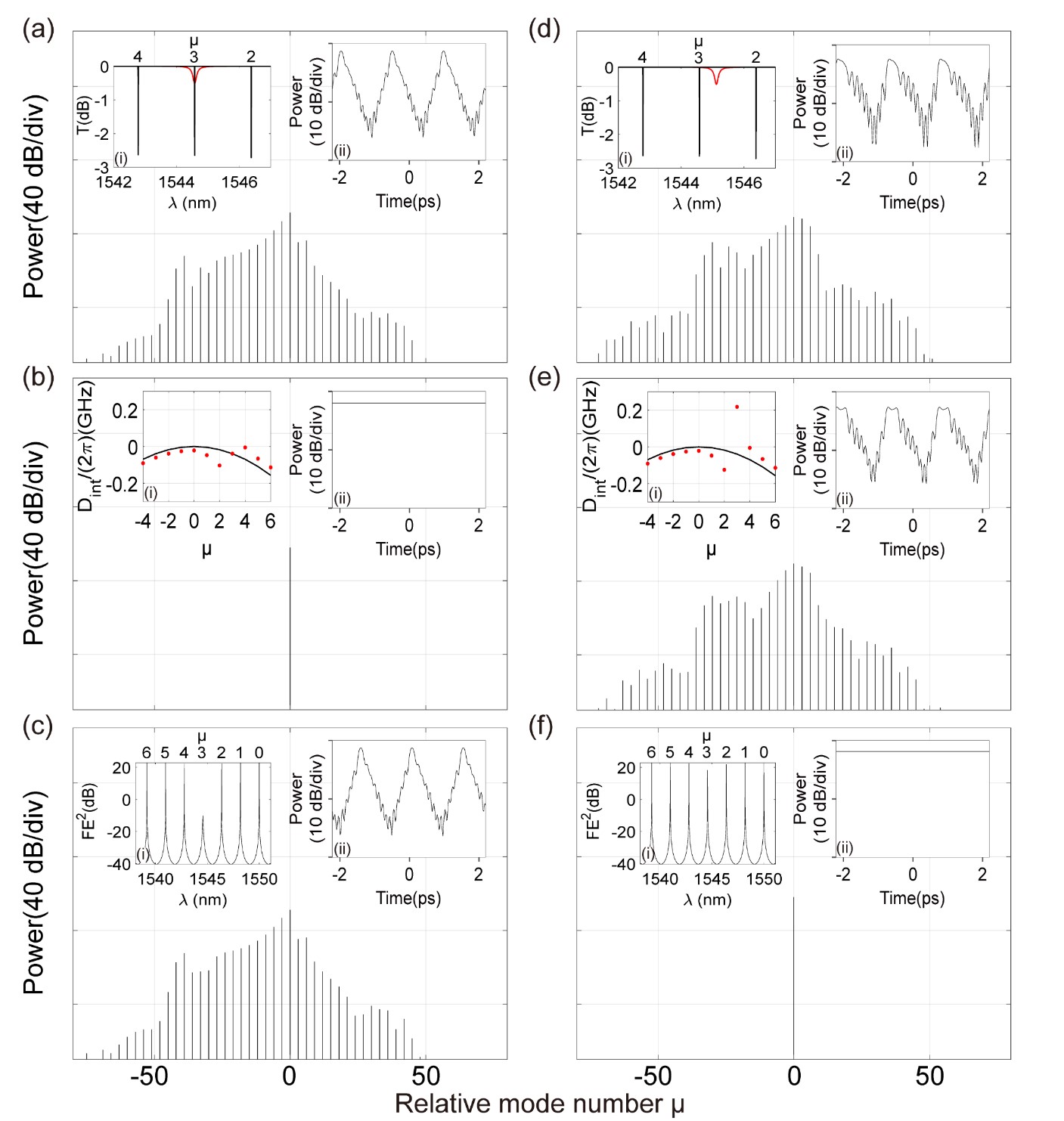}
    \caption{Numerical simulations of optical frequency combs and dark solitons generated at $\Delta=0$ (a)-(c) and $\Delta<0$ (d)-(f) in normal GVD regime based on GLLE with lumped filter $\tau_{st}(\omega)$. The filter functions used in (a),(d)/(b),(e)/(c),(f) correspond to $\tau_{st}(\omega)$, $arg(\tau_{st}(\omega))$ and $|\tau_{st}(\omega)|$, respectively. Insets in (a) and (d) on the left give the transmission spectra around $\mu=3$ by considering main and auxiliary cavity alone. The comb lines in all panels correspond to the intracavity power of the main cavity. Intracavity power of the auxiliary cavity are relatively weak, as discussed in Supplemental Material S5. Simulation parameters are as $t_{R_1}=4.4ps$, $L_1=628\mu$m, $\alpha_1=2.4\times10^{-3}$, $\alpha_2=1.5\times10^{-2}$, $u_1=0.0194$, $u_2=0.021$, $\beta_2=190ps^2/$km, $\Gamma=1\left(W\cdot m\right)^{-1}$, $\left|E_{in}\right|^2=0.42$W. $\delta$ scans linearly starting from -0.0041 at speed of $7.55\times{10}^{-7}$ per round-trip. $L_1/L_2$ is 33. See TABLE. S1 in Supplemental Material S2 for parameter description.}
    \label{p2}
\end{figure}

\textit{Dispersion-free single-mode depletion enabled MI and soliton formation in normal GVD regime.}
——MI, featured by the exponential growth under perturbations, can form first modulation sideband from noise and consequently generate optical frequency comb through four-wave mixing. In contrast to dispersion induced MI, depletion breaks the limit of Hermitian phase matching \cite{8868166,El-Ganainy:15} and leads to MI in normal GVD regime in fiber-based system. We now study MI generation in normal GVD microresonators under the single-mode depletion, as shown in Fig. 2(a). Since two cavities are aligned at $\mu=3$, mode at $\mu=3$ gets depleted, as confirmed by the inset of Fig. 2(c). Meanwhile, optical frequency comb is generated with FSR equals three times $D_1$ and three dark solitons are observed within a roundtrip time, indicating the coherence of frequency comb. Such nonlinear behavior is captured using a generalised version of standard Lugiato-Lefever Equation (GLLE) \cite{perego_gain_2018} where the impact of auxiliary ring is considered as a lumped spectral filter. The filter function, $\tau_{st}(\omega)$, is the transmission function of the subsystem shown by the dashed box in Fig. 1(b), i.e., $\tau_{st}(\omega)=\left(r_2-b_2e^{i\phi_2(\omega)}\right)/\left(1-b_2r_2e^{i\phi_2(\omega)}\right)$, where $r_2$, $b_2$ and $\phi_2(\omega)$ is the field coupling coefficient of two cavities, round trip transmission and round trip phase of the auxiliary cavity, respectively. Note that $\tau_{st}$($\omega$) uses the notations from power coupling model \cite{van_optical_2016}. A detailed comparison between power coupling model and energy coupling model from which Eq. (1) is derived is discussed in Supplemental Material S1. Verification of GLLE modelling is given in Supplemental Material S2. Obviously, $\tau_{st}(\omega)$ is a complex function and thus contains dispersion and dissipation effects, indicated by dispersive part of filter function $arg(\tau_{st}(\omega))$ and dissipative part of filter response $|\tau_{st}(\omega)|$, as further identified in Figs. 2(b) and 2(c), respectively. The remarkable coherence between Figs. 2(c) and 2(a) reveals the fact that dissipation is responsible for MI in microresonators in this scenario and produce coherent frequency comb that leads to dark solitons. Instead, the absence of sidebands in Fig. 2(b) is due to the fact that the all modes remains in the normal GVD regime. It is clear that the slight modification of dispersion at $\mu\neq3$ due to limited FSR has trivial impact in this case since no frequency comb is generated.

Figures 2(d)-2(f) show respectively the frequency and temporal features of the comb spectrum for full filter function $\tau_{st}(\omega)$, $arg(\tau_{st}(\omega))$ and $|\tau_{st}(\omega)|$, under the red-detuned scenario (blue-detuned scenario is examined in Supplemental Material S3). The local dispersion variation is responsible for comb generation, as evident from comparison of Figs. 2(e)-2(f) with Fig. 2(d). Indeed, the dispersion at $\mu=3$ becomes anomalous (rf. inset of Fig. 2(e)), which is consistent with the FSR of the generated comb as well as the number of dark solitons per round-trip time \cite{xue_mode-locked_2015,liu_investigation_2014}. In contrast, there is no frequency comb generation in Fig. 2(f), i.e., due to the impact of $|\tau_{st}(\omega)|$, since all resonant modes are effectively enhanced rather than dissipated. Depending on the relative dispersion and dissipation magnitude, we find that dispersion and dissipation can introduce MI simultaneously. Supplemental Material S4 studies an example that both dispersion and dissipation introduce MI but dissipation dominates, wherein FSR of the generated comb is determined by dissipation other than dispersion. Note that only intracavity powers of the main cavity are shown in Fig. 2, since optical fields in auxiliary ring are relatively weak and effective nonlinearities occur mostly in the main cavity (see discussion in Supplemental Material S5).
\begin{figure}[htbp]
    \centering
    \includegraphics[scale=0.52]{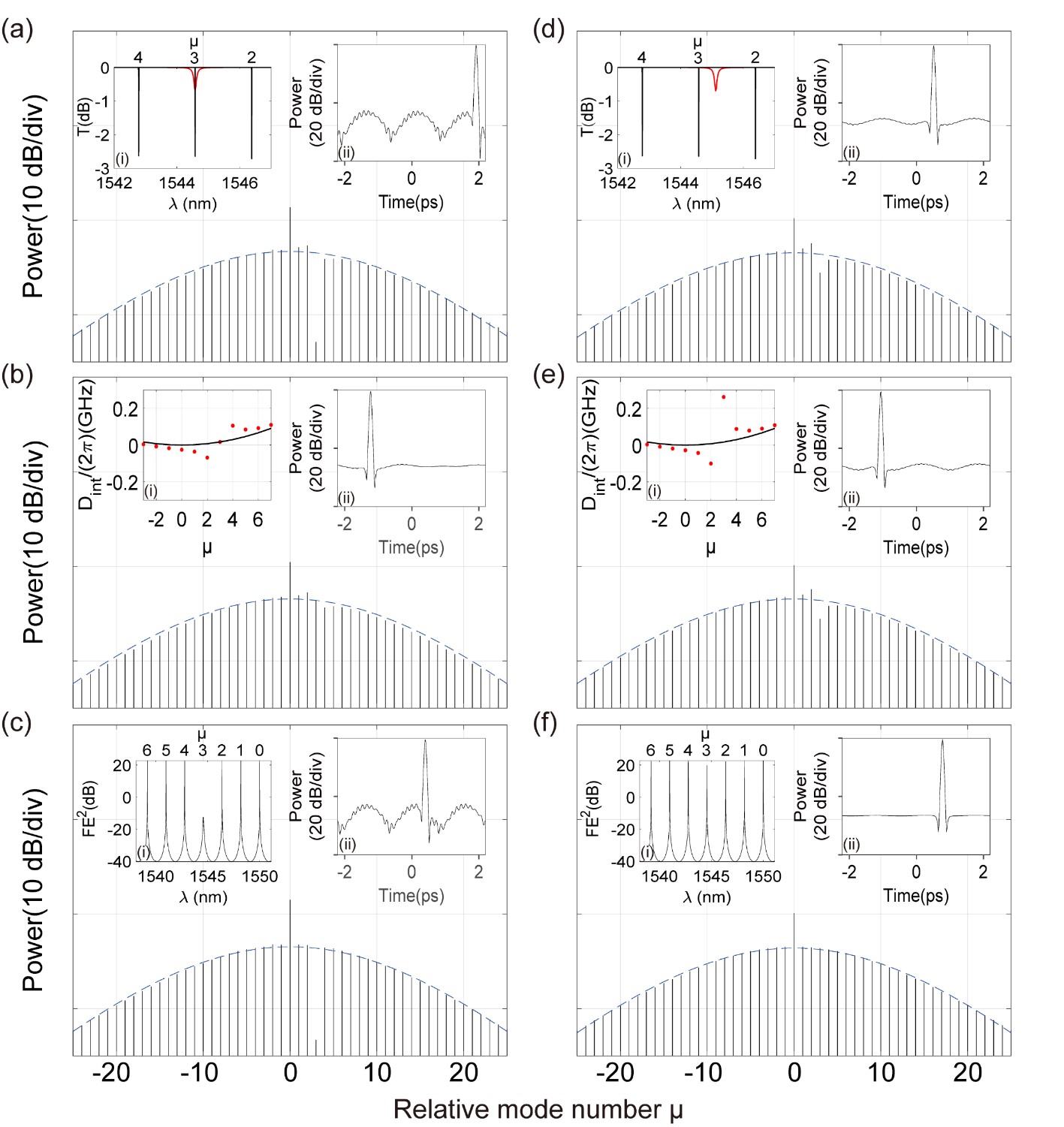}
    \caption{ Dissipative Kerr solitons generated at $\Delta=0$ (a)-(c) and $\Delta<0$ (d)-(f) in anomalous GVD regime based on GLLE with lumped filter $\tau_{st}(\omega)$. The filter functions used in (a),(d)/(b),(e)/(c),(f) correspond to $\tau_{st}(\omega)$, $arg(\tau_{st}(\omega))$ and $|\tau_{st}(\omega)|$, respectively (see Supplemental Material Fig. S2 for parameter details). Dashed line: $Sech^2$ envelop of single soliton case without depletion or mode-coupling induced dispersion. Simulation parameters are the same as Fig. 2 except $\alpha_2=1\times10^{-2}$, $u_2=0.02$ and $\beta_2=-81ps^2/km$.}
    \label{p3}
\end{figure}
\par
\textit{Single-mode depletion enhanced deterministic single soliton generation in anomalous GVD regime.}——The second application of our proposed selective single-mode depletion is deterministic single soliton generation in anomalous GVD regime, which usually is undermined by spontaneous synthesis of solitons from continuous-wave pump. Earlier attempt towards deterministic single soliton generation includes explorations of mode-coupling induced dispersive effect, auxiliary laser heating approach, forward and backward tuning technique and periodic optical pulses pump, fundamental-second-harmonic mode coupling \cite{bao_spatial_2017,liao_enhanced_2017,zhou_soliton_2019,guo_universal_2017,obrzud_temporal_2017,Xue:18}. Here we show that mode-selective dissipation in anomalous GVD region can promote deterministic single soliton generation as well. The comb spectra and temporal waveforms in case of $\Delta=0$ and $\Delta<0$ are shown in Figs. 3(a)-3(c) and 3(d)-3(f), respectively (see Supplemental Material S3 for $\Delta>0$ case). In comparison with the perfect $sech^2$ envelop of single soliton generated in a single microring resonator (blue dashed line), dissipation assisted single soliton generation creates a relative smooth envelop except a single comb line is absent ($\mu=3$ in this case), which is favorable in applications such as wavelength division multiplexing in fiber-optic communications (Fig. 3(a)). The residual dispersion induced spectrum modification of neighbouring modes respect to $sech^2$ envelop by our scheme is kept moderate, as confirmed by Fig. 3(b). More importantly, the probability of single soliton generation is found to be significantly improved under single mode depletion. In concrete, Fig. 4 shows the diagram of the probability of single soliton generation as a function of external pump power and pump phase detuning \cite{tikan_emergent_2021}, with and without single-mode depletion, showing greatly boosted success rate of single soliton generation to nearly 100\% even at low power levels. Single solitons are formed by fixed-speed unidirectional pump frequency sweeping in both cases, showing a simple way to greatly improved the success rate while avoiding complex pump scanning trajectory \cite{guo_universal_2017}. Mode at $\mu=10$ is depleted in this case, which is similar to the case of depleting $\mu=3$ or other mode located relatively close to the pump. The improvement in single soliton statistics can be attributed to the temporal modulation induced by spectral filtering. The difference of single soliton existence boundary in two cases needs further stability analysis of the GLLE \cite{godey_stability_2014} which is out of the scope of this paper.
\begin{figure}[htbp]
    \centering
    \includegraphics[scale=0.32]{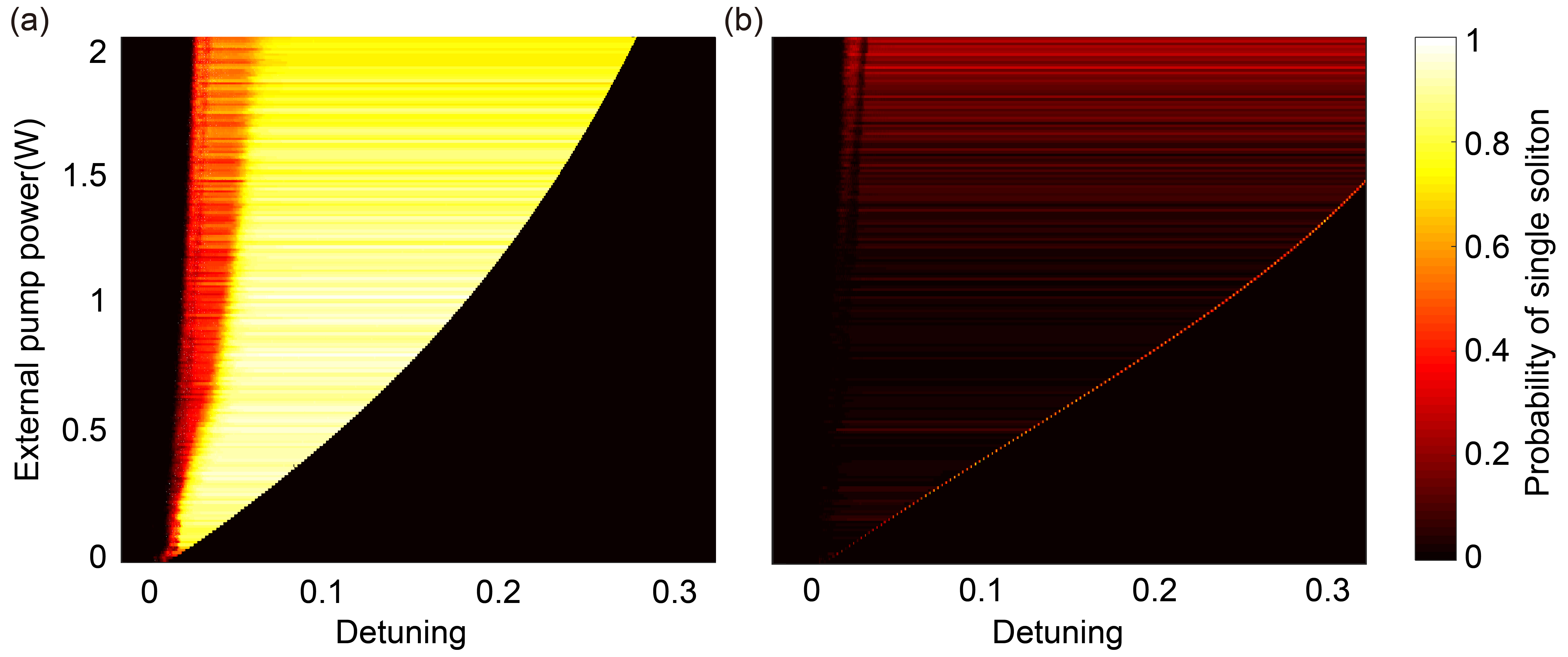}
    \caption{Probability of single soliton formation on the pump power and detuning map with (a) and without (b) single-mode depletion in anomalous GVD regime. The mode $\mu=10$ is depleted. Parameters used are the same as Fig. 3(a).}
    \label{p4}
\end{figure}

In conclusion, we studied selective single mode depletion effect between two coupled microring resonators in the context of nonlinear optical process. As the resonant frequency of the high-Q (main) ring resonator is aligned with that of a lossy resonator, with properly tuned yet experimentally feasible parameters, the relative field-amplitude of the main resonator is completely suppressed, together with the absence of resonance split (dispersion-free). With tuneable resonant frequency of the lossy resonator, one thus is able to selectively deplete any target resonance of the main ring resonators against the broad background comb spectra, leading to microcombs with reconfigurable FSR \cite{xue_normal-dispersion_2015}. Such selective and dispersion-free singe mode depletion essentially relies on the simultaneous collapse of eigenvalues and eigenvectors of the coupled microring resonators at exceptional point, wherein the impact of dispersion is effectively disentangled from dissipation. Importantly, selective and dispersion-free singe mode depletion is particularly relevant to dissipative Kerr solitons in high-Q microresonators, as the hidden MI and deterministic single soliton generation enabled by pure mode depletion are disclosed explicitly for the first time. We expect that the conception revealed in this paper trigger wide exploration of non-Hermitian control of micro-cavities in an upgraded parameter space and have broad impact on a wide optics field including optical signal processing, quantum optics, ultrafast optics etc.
\bibliography{output.bib}

\end{document}